\documentclass[twocolumn,superscriptaddress,nobalancelastpage,10pt,pra,a4paper]{revtex4}
\usepackage{amsmath}
\usepackage{amsthm}
\usepackage{amssymb}
\usepackage{amsfonts}
\usepackage{graphicx}
\usepackage{wasysym}
\usepackage{mathrsfs}
\usepackage{yfonts}
\usepackage{bbold}
\usepackage{verbatim}
\usepackage{subfigure}
\usepackage{color}

\newcommand{\ket}[1]{\left| #1 \right\rangle}
\newcommand{\bra}[1]{\left\langle #1 \right|}

\makeatletter
\newcommand*{\rom}[1]{\expandafter\@slowromancap\romannumeral #1@}
\makeatother

\makeatletter
\newcommand{\roml}[1]{\lowercase\expandafter{\romannumeral #1\relax}}
\makeatother

\DeclareMathAlphabet{\mathpzc}{OT1}{pzc}{m}{it}

\begin{document}


\title{Many-Particle Interferometry and Entanglement by Path Identity}

\author{Mayukh Lahiri}
\email{mayukh.lahiri@univie.ac.at} \affiliation{Faculty of Physics,
University of Vienna, Boltzmanngasse 5, Vienna A-1090,
Austria.}\affiliation{Institute for Quantum Optics and Quantum
Information, Austrian Academy of Sciences, Boltzmanngasse 3, Vienna
A-1090, Austria.}

\begin{abstract}
We introduce a general scheme of many-particle interferometry in
which two identical sources are used and ``which-way information''
is eliminated by making the paths of one or more particles identical
(path identity). The scheme allows us to generate many-particle
entangled states. We provide general forms of these states and show
that they can be expressed as superpositions of various Dicke
states. We illustrate cases in which the scheme produces maximally
entangled two-qubit states (Bell states) and maximally three-tangled
states (three-particle Greenberger-Horne-Zeilinger-class states). A
striking feature of the scheme is that the entangled states can be
manipulated without interacting with the entangled particles; for
example, it is possible to switch between two distinct Bell states.
Furthermore, each entangled state corresponds to a set of
many-particle interference patterns. The visibility of these
patterns and the amount of entanglement in a quantum state are
connected to each other. The scheme also allows us to change the
visibility and the amount of entanglement without interacting with
the entangled particles and, therefore, has the potential to play an
important role in quantum information science.
\end{abstract}

\maketitle

\begin{figure}[htbp]  \centering
\includegraphics[width=0.3\textwidth]{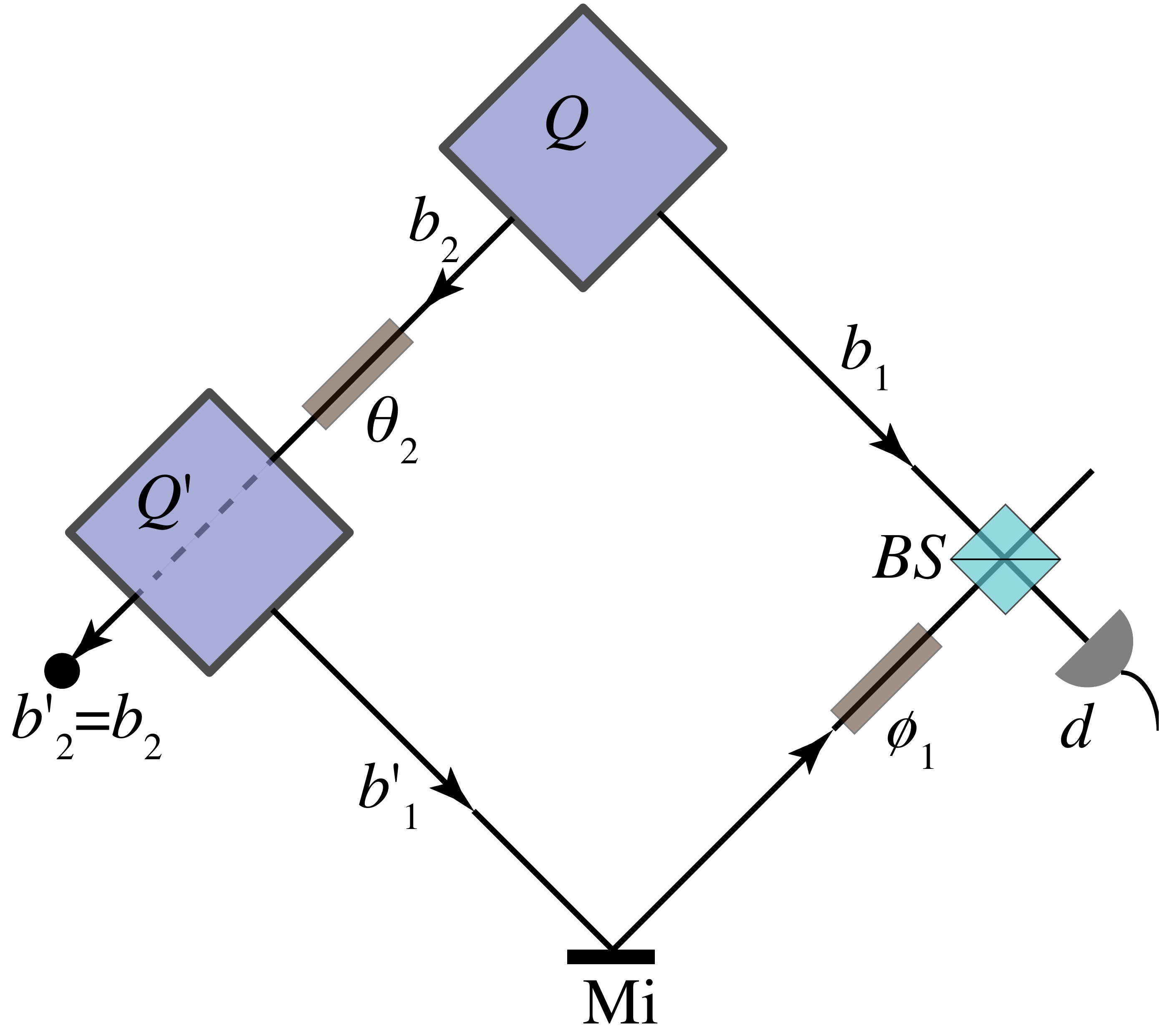}
\qquad \caption{Zou-Wang-Mandel experiment. $Q$ and $Q'$ are two
identical sources emitting two photons ($1,2$) into beams
($b_1$,$b_2$) and ($b_1'$,$b_2'$). The paths of photon 2 are made
identical by sending the beam $b_2$ through $Q'$ and aligning it
with $b_2'$. Photon 2 is not detected. Single-photon interference is
observed at the detector $d$ when $b_1$ and $b_1'$ are superposed.}
\label{fig:Mand-setup}
\end{figure}

\emph{Introduction}.\textemdash In 1991, Zou, Wang, and Mandel
reported observation of single-photon interference by using two
identical two-photon sources \cite{ZWM-ind-coh-PRL,WZM-ind-coh-PRA}.
A striking feature of their experiment, which was originally
suggested by Ou, was to make the paths of the same photon generated
by the two sources identical (Fig. \ref{fig:Mand-setup}). This path
identity created coherence between the beams ($b_1$ and $b_2$) of
the other photon and a single-photon pattern resulted. The
interference pattern could be manipulated without interacting with
the photon that was detected. In a recent series of work the concept
of path identity has been applied to imaging
\cite{LBCRLZ-mandel-im,LLLZ-th-mandel-img}, spectroscopy
\cite{Kulik-spec-2016}, generating a light beam in any state of
polarization \cite{LHLBZ-quant-pol}, fundamental test of quantum
mechanics \cite{HMM-comp-two-ph-PRL,HMM-comp-two-ph-PRA}, measuring
correlations between two photons
\cite{HLLLZ-mom-corr-exp,LHLLZ-mom-corr-exp}, and generating
multiphoton high-dimensional entangled states
\cite{ent-path-id-PRL-2017}.
\par
The aim of this paper is to introduce a general scheme of generating
many-particle entangled states and many-particle interference
patterns by applying the method of path identity. An important
feature of this scheme is that the generated entangled states (and
also the interference patterns) can be manipulated without
interacting with the entangled particles.
\par
For the sake of clarity, we begin by discussing two special cases
and then introduce the scheme in its most general form.
\par
\begin{figure}\centering
 \subfigure[] {
    \label{figa:3-part-1-Pi-setup}
    \includegraphics[width=0.35\textwidth]{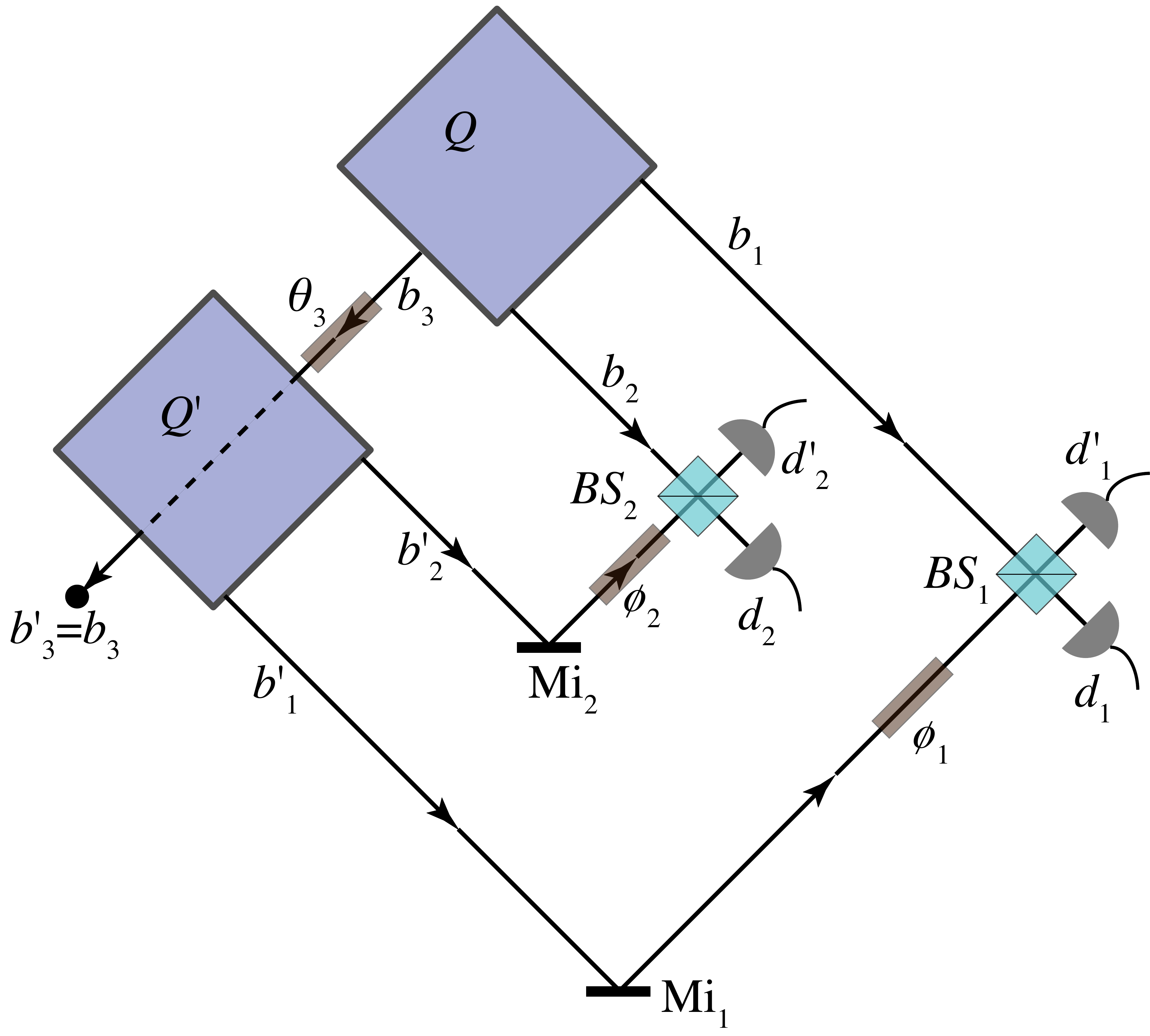}
} 
   \subfigure[] {
    \label{figb:3-part-1-Pi-patterns}
    \includegraphics[width=0.3\textwidth]{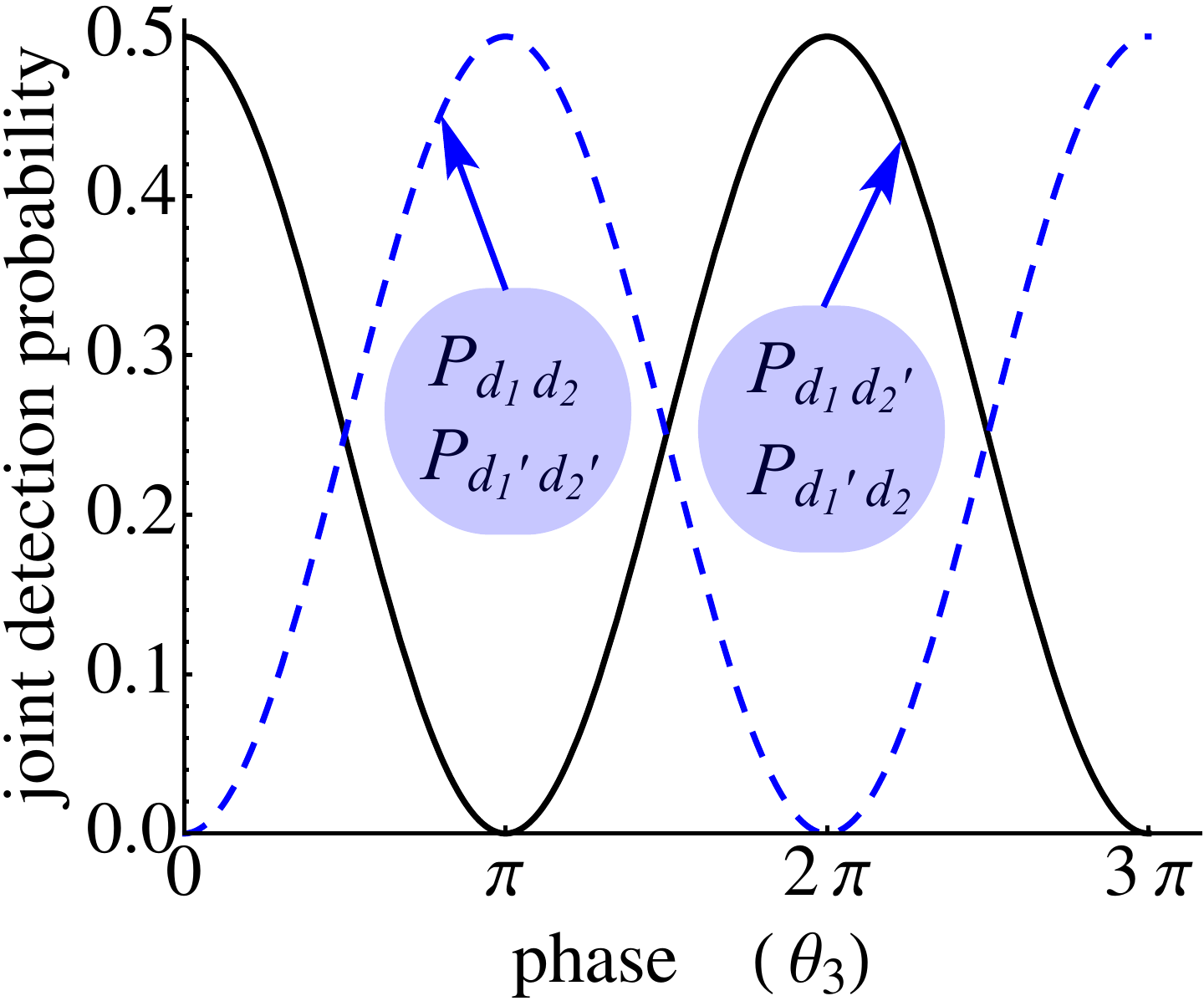}
} 
     \caption{Two-particle interference and entanglement by one-particle path identity: (a) Schematic of the setup. $Q$ and $Q'$ are
two identical three-particle sources emitting particles ($1$,$2$,$3$) into beams ($b_1$,$b_2$,$b_3$) and ($b_1'$,$b_2'$,$b_3'$).
Beam $b_3$ is aligned with $b_3'$ in such a way that it is not
possible to determine the source of the particle $3$ if observed
after $Q'$. The phase change along $b_3$ due to propagation from $Q$
to $Q'$ is $\theta_3$. Beams $b_1$ and $b_1'$ are superposed by BS1
(beam splitter or an equivalent device) with two outputs detected at
$d_1$ and $d_1'$. The phase difference between $b_1$ and $b_1'$ is
$\phi_1$. Likewise $b_2$ and $b_2'$ are superposed by BS2 with two
outputs at $d_2$ and $d_2'$; the corresponding phase difference is
$\phi_2$. Particle $3$ is never detected. (b) Two-particle
interference patterns. Probabilities ($P_{d_1d_2}$, $P_{d_1d_2'}$,
$P_{d_1'd_2}$, $P_{d_1'd_2'}$) of joint detection at the pairs of
detectors ($d_1$,$d_2$), ($d_1$,$d_2'$), ($d_1'$,$d_2$), and
($d_1'$,$d_2'$) vary sinusoidally with the phase $\theta_3$ that can
only be modulated using particle $3$. Interference patterns
$P_{d_1d_2}$ and $P_{d_1'd_2'}$ are in phase (dashed line). These
patterns are complementary to the patterns $P_{d_1d_2'}$ and
$P_{d_1'd_2}$ (solid line). We have set $\Phi^{(2)}=2n\pi$, $n$
being an integer. A maximum and a minimum of an interference pattern
are attained for two distinct Bell states.} \label{fig:3-part-1-Pi}
\end{figure}

\emph{Case \rom{1}} (Fig. \ref{fig:3-part-1-Pi}).\textemdash Suppose that
a three-particle source, $Q$, emits particles $1$, $2$, and $3$ into
the beams $b_1$, $b_2$, and $b_3$, respectively [Fig.
\ref{figa:3-part-1-Pi-setup}]. We now consider another identical
source, $Q'$, whose emitted beams are denoted by $b_1'$, $b_2'$, and
$b_3'$. If the two sources emit in quantum superposition \cite{Note-q-sup}, the
three-particle state is given by
\begin{align}\label{state-thr-part}
\ket{X_3}=
(\ket{b_1}_1\ket{b_2}_2\ket{b_3}_3+e^{i\phi_0}\ket{b_1'}_1\ket{b_2'}_2\ket{b_3'}_3)/\sqrt{2},
\end{align}
where $\ket{b_1}_1$ denotes particle $1$ in beam $b_1$, etc.,
$\phi_0$ is a phase factor, and we have assumed that emission
probability at the two sources are equal. Note that $\ket{X_3}$ is a
three particle Greenberger-Horne-Zeilinger (GHZ) state
\cite{GHZ-original,Z-three-part-interf}.
\par
Suppose now that the paths of particle $3$ emitted by $Q$ and $Q'$
are made identical ($b_3=b_3'$). This can be done by sending beam
$b_3$ through $Q'$ and aligning it with $b_3'$ [Fig.
\ref{figa:3-part-1-Pi-setup}]. We therefore have $\ket{b_3}_3 \to
\exp[i\theta_3]\ket{b_3'}_3,$ where $\theta_3$ can be interpreted as
the phase gained due to propagation from $Q$ to $Q'$. Applying this
transformation to Eq. (\ref{state-thr-part}), we find that $\ket{X}
\to \ket{\psi_0}$, where \cite{Note-qft}
\begin{align}\label{state-thr-part-align}
\ket{\psi_0}=\frac{1}{\sqrt{2}}
(\ket{b_1}_1\ket{b_2}_2+e^{i(\phi_0-\theta_3)}\ket{b_1'}_1\ket{b_2'}_2)
\ket{b_3'}_3 .
\end{align}
This state is a tensor product of a ``spin-free'' two-particle
entangled state \cite{Note-spin-free-states} and a single third particle state.
\par
The beams $b_1$ and $b_1'$ are superposed by a 50-50 beam splitter
(or an equivalent device), BS1, and the two outputs are received by
detectors $d_1$ and $d_1'$. The phase difference between the beams
$b_1$ and $b_1'$ is given by $\phi_1$. Likewise  $b_2$ and $b_2'$
are superposed (corresponding phase difference $\phi_2$) by BS2 with
outputs at $d_2$ and $d_2'$. The consequent transformations of the
pairs of kets are therefore given by
\begin{subequations}\label{BS-1}
\begin{align}
&\ket{b_j}_j \to (\ket{d_j}_j+i\ket{d_j'}_j)/\sqrt{2}, \label{BS-1-1} \\
&\ket{b_j'}_j \to e^{i\phi_j} (\ket{d_j'}_j+i\ket{d_j}_j)/\sqrt{2},
\label{BS-1-2}
\end{align}
\end{subequations}
where $j=1,2$. Applying the evolution given by Eq. (\ref{BS-1}) to
the state in Eq. (\ref{state-thr-part-align}), we find that
\begin{align}\label{state-thr-part-final}
&\ket{\psi_0} \to
\ket{\psi}=\frac{1}{2}\Big\{(1-e^{i\zeta^{(3)}_1})\frac{1}{\sqrt{2}}
(\ket{d_1}_1\ket{d_2}_2-\ket{d_1'}_1\ket{d_2'}_2) \nonumber
\\ & +i(1+e^{i\zeta^{(3)}_1})\frac{1}{\sqrt{2}}
(\ket{d_1}_1\ket{d_2'}_2+\ket{d_1'}_1\ket{d_2}_2) \Big\}\ket{b'_3}_3.
\end{align}
where $\zeta^{(3)}_1=\phi_0+\phi_1+\phi_2-\theta_3$. The complex coefficients associated with
$\ket{d_1}_1\ket{d_2}_2$, $\ket{d_1}_1\ket{d_2'}_2$,
$\ket{d_1'}_1\ket{d_2}_2$, and $\ket{d_1'}_1\ket{d_2'}_2$ are the
probability amplitudes of joint (coincidence) detection of particles
$1$ and $2$ at the pairs of detectors ($d_1$,$d_2$), ($d_1$,$d_2'$),
($d_1'$,$d_2$), and ($d_1'$,$d_2'$), respectively. The coincidence
detection rate at these pairs of detectors are given by the
corresponding probabilities (square of the modulus of the
probability amplitudes), i.e., by
\begin{subequations}\label{two-part-pattern}
\begin{align}
P_{d_1d_2}=P_{d_1'd_2'}&=\frac{1}{4}[1-\cos(\Phi^{(2)}-\theta_3)], \label{two-part-pattern:a} \\
P_{d_1d_2'}=P_{d_1'd_2}&=\frac{1}{4}[1+\cos(\Phi^{(2)}-\theta_3)],
\label{two-part-pattern:b}
\end{align}
\end{subequations}
where $\Phi^{(2)}=\phi_0+\phi_1+\phi_2$, i.e., $\zeta^{(3)}_1=\Phi^{(2)}-\theta_3$.
\par
Clearly, two-particle interference
\cite{GM-two-ph-interf,HOM-effect,Shi-2pinterf-Bell,Z-two-part-interf}
involving $1$ and $2$ will occur. The fact that $\ket{b'_3}_3$ gets
factored out in Eq. (\ref{state-thr-part-final}) implies that one
does not need to detect particle $3$ to observe the interference of
$1$ and $2$. However, the two-particle interference patterns can be
modulated by using this undetected particle [Fig.
\ref{figb:3-part-1-Pi-patterns}], as is evident from the appearance
of $\theta_3$ in the joint-detection probabilities. Equation
(\ref{two-part-pattern}) shows that the two-particle interference
patterns at the two pairs of detectors ($d_1$,$d_2$) and
($d_1'$,$d_2'$) are identical. Similarly, the patterns observed at
($d_1'$,$d_2$) and ($d_1$,$d_2'$) are also identical. The patterns
observed in the former set of detector pairs are complementary to
those observed in the latter set of detector pairs [Fig.
\ref{figb:3-part-1-Pi-patterns}].
\par
We now note that the pair of
particles (1,2) will be the following two distinct Bell states for
$\zeta^{(3)}_1=2m\pi$ and $\zeta^{(3)}_1=(2m+1)\pi$, respectively:
\begin{subequations}\label{Bell-st-output}
\begin{align}
\ket{\Psi^{+}}&=\frac{1}{\sqrt{2}}
(\ket{d_1}_1\ket{d_2'}_2+\ket{d_1'}_1\ket{d_2}_2), \label{Bell-st-output:a} \\
\ket{\Phi^{-}}&=\frac{1}{\sqrt{2}}
(\ket{d_1}_1\ket{d_2}_2-\ket{d_1'}_1\ket{d_2'}_2),
\label{Bell-st-output:b}
\end{align}
\end{subequations}
where $m=0,\pm 1,\pm 2,\dots$. A comparison between Eqs.
(\ref{two-part-pattern}) and (\ref{Bell-st-output}) shows that when
the state $\ket{\Psi^{+}}$ is obtained, the coincidence counts at
($d_1'$,$d_2$) and ($d_1$,$d_2'$) maximize and the coincidence
counts at ($d_1$,$d_2$) and ($d_1'$,$d_2'$) minimize [see Fig.
\ref{figb:3-part-1-Pi-patterns}]. Likewise, the state
$\ket{\Phi^{-}}$ is obtained when coincidence counts are maximum at
($d_1$,$d_2$) and ($d_1'$,$d_2'$), and minimum at ($d_1'$,$d_2$) and
($d_1$,$d_2'$). \emph{The system therefore allows one to switch
between the two Bell states without any interaction with the pair of
particles.}
\begin{figure}[htbp]  \centering
\includegraphics[width=0.4\textwidth]{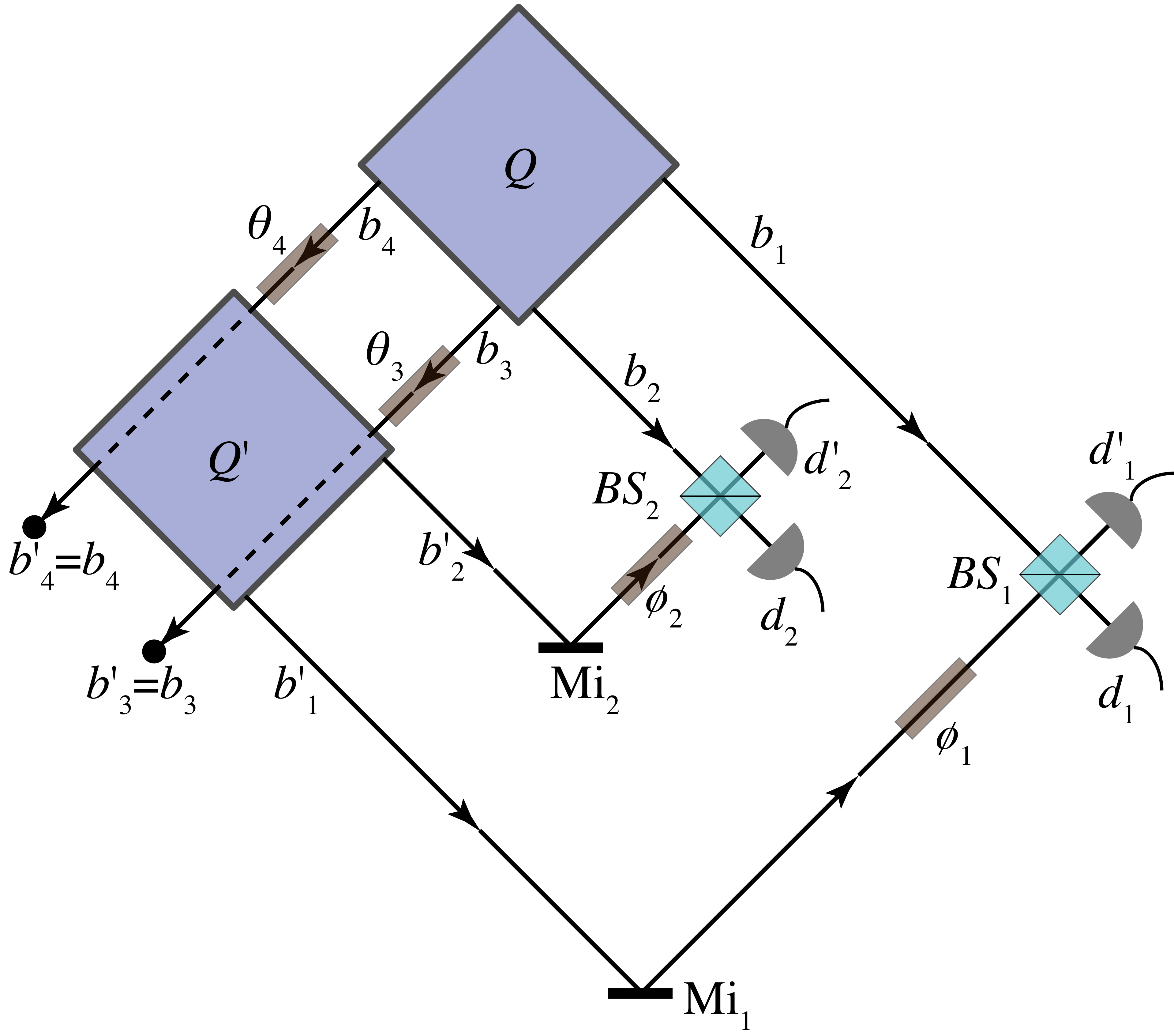}
\qquad \caption{Two-particle interference and entanglement by
two-particle path identity: $Q$ and $Q'$ are two identical
four-particle sources emitting particles ($1$,$2$,$3$,$4$) into
beams ($b_1$,$b_2$,$b_3$,$b_4$) and ($b_1'$,$b_2'$,$b_3'$,$b_4'$),
respectively. The beams $b_3$ and $b_4$ are aligned with $b_3'$ and
$b_4'$, respectively; the corresponding phase changes are $\theta_3$
and $\theta_4$. Particles $3$ and $4$ are not detected. The rest of
the notations are same as in Fig. \ref{figa:3-part-1-Pi-setup}. The
two-particle interference patterns produced in this setup are
identical to those shown in Fig. \ref{fig:3-part-1-Pi}, except each
pattern can now be modulated by both $\theta_3$ and $\theta_4$. The
same Bell states are obtained under conditions strictly similar to
Case \emph{\rom{1}}.} \label{fig:4-part-2Pi}
\end{figure}
\par
\emph{Case \rom{2}} (Fig. \ref{fig:4-part-2Pi}).\textemdash We now
consider two four-particle sources, $Q$ and $Q'$, emitting in
quantum superposition. $Q$ and $Q'$ emit particles $1$, $2$, $3$,
and $4$ into the beams $(b_1,b_2,b_3,b_4)$ and
$(b_1',b_2',b_3',b_4')$, respectively (Fig. \ref{fig:4-part-2Pi}).
The resulting quantum state is given by
\begin{align}\label{state-four-part}
\ket{X_4}=\frac{1}{\sqrt{2}}(\prod_{j=1}^4
\ket{b_j}_j+e^{i\phi_0}\prod_{j=1}^4\ket{b_j'}_j).
\end{align}
Beams $b_3$ and $b_4$ are sent through $Q'$ and are perfectly
aligned with beams $b_3'$ and $b_4'$ (path identity). The
corresponding transformations of kets are given by $\ket{b_3}_3 \to
\exp[i\theta_3]\ket{b_3'}_3$ and $\ket{b_4}_4 \to
\exp[i\theta_4]\ket{b_4'}_4$. The beams of particles $1$ and $2$ are
superposed in the same way as in case \emph{\rom{1}}. Following
theoretical steps which are strictly similar to case \emph{\rom{1}},
we find that the two-particle interference patterns are given by
\begin{subequations}\label{two-part-pattern-2}
\begin{align}
P_{d_1d_2}=P_{d_1'd_2'}&=\frac{1}{4}[1-\cos(\Phi^{(2)}-\theta_3-\theta_4)], \label{two-part-pattern-2:a} \\
P_{d_1d_2'}=P_{d_1'd_2}&=\frac{1}{4}[1+\cos(\Phi^{(2)}-\theta_3-\theta_4)],
\label{two-part-pattern-2:b}
\end{align}
\end{subequations}
where $\Phi^{(2)}$ is defined below Eq. (\ref{two-part-pattern}).
\par
Let us define $\zeta^{(4)}_2\equiv \Phi^{(2)}-\theta_3-\theta_4$. It
again follows that the pair of particles, ($1,2$), will be in the
Bell states given by Eqs. (\ref{Bell-st-output:a}) and
(\ref{Bell-st-output:b}) for $\zeta^{(4)}_2=2m\pi$ and
$\zeta^{(4)}_2=(2m+1)\pi$, respectively.
\par
Before introducing the general scheme, we compare cases \emph{\rom{1}} and \emph{\rom{2}} and
note the following: 1) the difference between the number of
particles produced by a source and the number of particles used for
path identity is the same; 2) both setups produce the same entangled
states; 3) an entangled state is obtained only when a maximum occurs
in a set of interference patterns; and 4) the entangled states and
the interference patterns can be modified without interacting with
the associated particles.
\begin{figure}[htbp]  \centering
\includegraphics[width=0.4\textwidth]{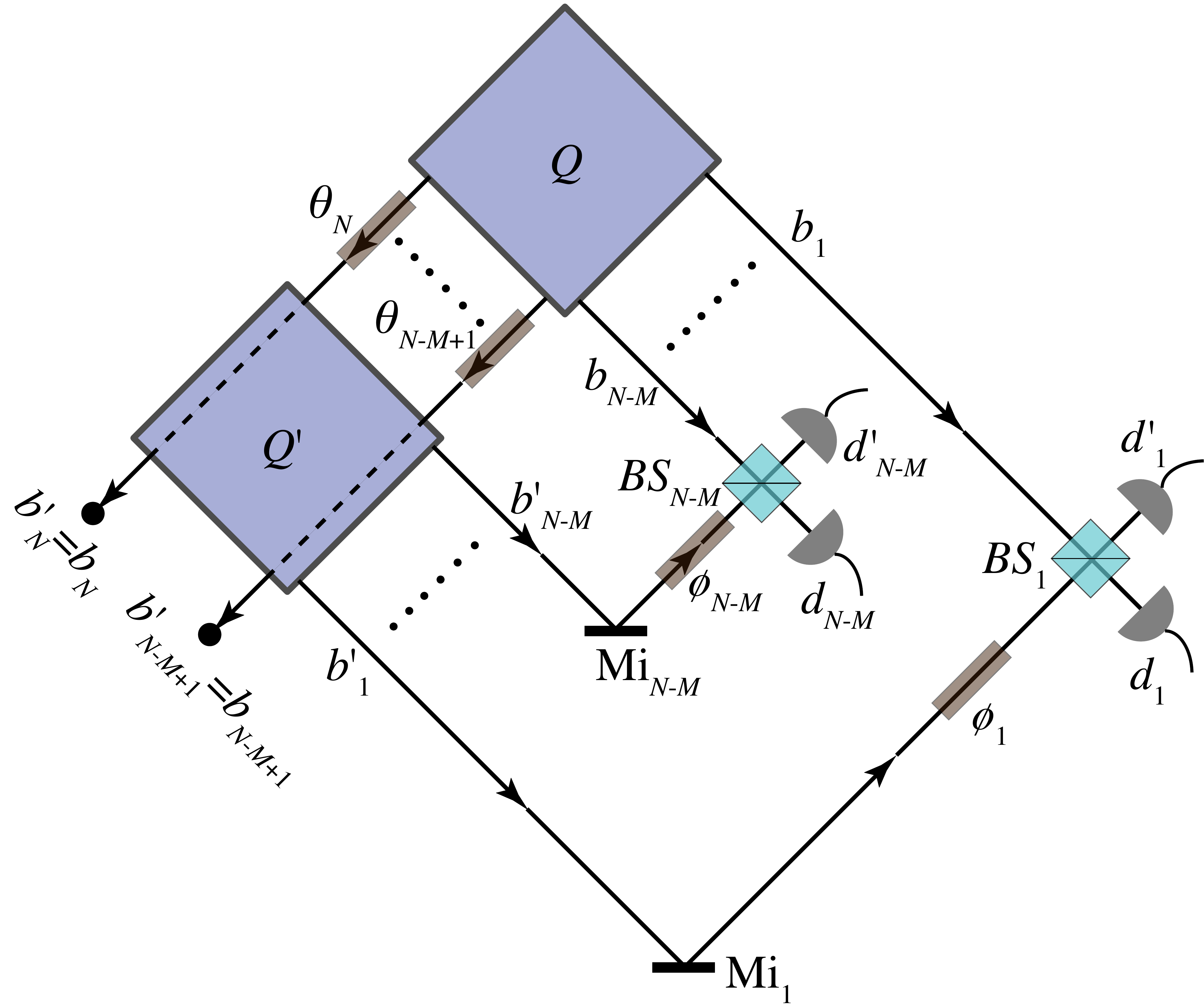}
\qquad \caption{The general scheme (notations are analogous to Figs.
2 and 3). Two identical $N$-particle sources emit particles
($1,2,\dots, N$) into beams ($b_1,b_2, \dots, b_N$) and ($b_1',b_2',
\dots, b_N'$), respectively. Paths of $M$ particles ($N-M+1,\dots,
N$) are made identical by aligning the corresponding beams and these
particles are not detected. Rest of the particles ($1,2,\dots, N-M$)
produce $N-M$-particle interference patterns and entangled states
when the corresponding beams are superposed.} \label{fig:gen-setup}
\end{figure}
\par
\emph{General Scheme} (Fig. \ref{fig:gen-setup}).\textemdash Let us
consider two identical sources, $Q$ and $Q'$, each of which can emit
$N$ particles, ($1,2 \dots, N$), into beams ($b_1,b_2, \dots, b_N$)
and ($b_1',b_2', \dots, b_N'$), respectively. The sources emit in
quantum superposition and thus produce the state
\begin{align}\label{state-N-part}
\ket{X_N}=\frac{1}{\sqrt{2}}(\prod_{j=1}^N
\ket{b_j}_j+e^{i\phi_0}\prod_{j=1}^N \ket{b_j'}_j).
\end{align}
Paths of the particles $N-M+1,\dots,N$ are made identical by sending
the beams $b_{N-M+1},\dots,b_{N}$ through $Q'$ and perfectly
aligning them with $b_{N-M+1}',\dots,b_{N}'$. These alignments lead to the set
of transformations
\begin{equation}\label{alignment-cond-gen}
\ket{b_l}_l \to \exp[i\theta_l]\ket{b_l'}_l, \quad l=N-M+1,\dots,N,
\end{equation}
where $\theta_l$ is the phase gained due to propagation from $Q$ to
$Q'$ along $b_l$. The pairs of beams ($b_1,b_1'$), ($b_2,b_2'$),
$\dots$, ($b_{N-M},b_{N-M}'$) are superposed by $N-M$ beam
splitters, $BS_1$, $BS_2$, $\dots$, $BS_{N-M}$. The outputs of the
beam splitters are detected at the pairs of detectors ($d_1,d_1'$),
($d_2,d_2'$), $\dots$, ($d_{N-M},d_{N-M}'$). The corresponding
transformations of kets are given by Eq. (\ref{BS-1}) with
$j=1,2,\dots, N-M$. We measure $(N-M)$-fold coincidences at a set of
$N-M$ detectors, each placed at an output of a distinct beam
splitter; an example of a set of detectors is ($d_1,d_2, \dots,
d_{N-M}$).
\par
Applying transformations (\ref{BS-1}) and (\ref{alignment-cond-gen})
to Eq. (\ref{state-N-part}), we find that the quantum states becomes
\begin{align}\label{state-N-part-final}
&\ket{\psi_N} \nonumber
\\ &=\left(\frac{1}{\sqrt{2}}\right)^{N-M+1}\left[\sum_{r=0}^{N-M}
(i^r+i^{N-M-r}e^{i\xi^{(N)}_M}) \ket{D_r}^{N-M}\right]
 \nonumber
\\ & \qquad \qquad \qquad \qquad \otimes \prod_{j=1}^M \ket{b_{N-M+j}'}_{N-M+j},
\end{align}
where
$\xi^{(N)}_M=\phi_0+\sum_{k=1}^{N-M}\phi_k-\sum_{j=1}^{M}\theta_{N-M+j}$
and the ($N-M$)-particle state, $\ket{D_r}^{N-M}$, is a Dicke state
\cite{Dicke-st-orig,Toth-Dickest-JOSAB}, i.e., a sum of
$\binom{N-M}{r}$ terms (states), each being a product of $r$ primed
states ($\ket{d_k'}_k$) and $N-M-r$ unprimed states ($\ket{d_k}_k$);
in our notation, $\ket{D_0}^{N-M}=\prod_{k=1}^{N-M}\ket{d_k}_k$ and
$\ket{D_{N-M}}^{N-M}=\prod_{k=1}^{N-M}\ket{d_k'}_k$ have one term
each.
\par
It follows from Eq. (\ref{state-N-part-final}) that when $N-M\geq
1$, the system produced $(N-M)$-particle interference patterns. The
fact that the states $\ket{b_{N-M+j}'}_{N-M+j}$ factor out implies
that in order to observe these patterns one does not need to detect
the $M$ particles used for path identity.
\par
The $N-M$ particles emerging from the outputs of the beam splitters
will be in different entangled states depending on the value of
$\xi^{(N)}_M$. One can express these states in simplified forms by
considering the cases $N-M=4n, 4n+1, 4n+2, 4n+3$, where $n=0, 1, 2,
\dots$. It follows from Eq. (\ref{state-N-part-final}) that the
forms are (dropping the normalization constant
$(1/\sqrt{2})^{N-M-1}$):\\ \roml{1}) for $N-M=4n>0$,
$\xi^{(N)}_M=2m\pi$; and $N-M=4n+2$, $\xi^{(N)}_M=(2m+1)\pi$:
\begin{align}\label{ent-state-gen-1}
\ket{F_1}=\sum_{r'=0}^{(N-M)/2} (-1)^{r'}\ket{D_{2r'}}^{N-M},
\end{align}
\roml{2}) for $N-M=4n>0$, $\xi^{(N)}_M=(2m+1)\pi$; and $N-M=4n+2$,
$\xi^{(N)}_M=2m\pi$:
\begin{align}\label{ent-state-gen-2}
\ket{F_2}=\sum_{r'=0}^{(N-M-2)/2} (-1)^{r'}\ket{D_{2r'+1}}^{N-M},
\end{align}
\roml{3}) for $N-M=4n+1$, $\xi^{(N)}_M=(2m-1/2)\pi$; and $N-M=4n+3$,
$\xi^{(N)}_M=(2m+1/2)\pi$:
\begin{align}\label{ent-state-gen-3}
\ket{F_3}=\sum_{r'=0}^{(N-M-1)/2} (-1)^{r'}\ket{D_{2r'}}^{N-M},
\end{align}
and \roml{4}) for $N-M=4n+1$, $\xi^{(N)}_M=(2m+1/2)\pi$; and
$N-M=4n+3$, $\xi^{(N)}_M=(2m-1/2)\pi$:
\begin{align}\label{ent-state-gen-4}
\ket{F_4}=\sum_{r'=0}^{(N-M-1)/2} (-1)^{r'}\ket{D_{2r'+1}}^{N-M},
\end{align}
where $m=0,\pm 1,\pm 2,\dots$. These entangled states [Eqs.
(\ref{ent-state-gen-1})-(\ref{ent-state-gen-4})] depend on the
difference $N-M$, not on individual values of $N$ and $M$.
\par
It is important to note the particles emerging from the beam
splitters can be transformed from one entangled state to other by
changing the phase $\xi^{(N)}_M$. Since $\xi^{(N)}_M$ contains the
phases $\theta_{N-M+j}$, it can be varied without interacting with
the entangled particles. Therefore, the scheme allows us to modify a
many-particle entangled state in an interaction-free way.
Furthermore, each of these states is generated when a maximum occurs
in a corresponding set of many-particle interference patterns. We
made these observations in the special cases \emph{\rom{1}} and
\emph{\rom{2}} discussed above.
\par
\emph{Case \rom{3}: GHZ-Class State}.\textemdash As another example
let us consider the case in which $N-M=3$. It follows from Eq.
(\ref{ent-state-gen-3}) that the system produces the states of the
form (replacing the unprimed states by $0$ and primed states by 1)
\begin{align}\label{GHZ-output}
\frac{1}{2}(&\ket{0}_{1}\ket{0}_{2}\ket{0}_{3}-\ket{1}_{1}\ket{1}_{2}\ket{0}_{3}-
\ket{1}_{1}\ket{0}_{2}\ket{1}_{3}\nonumber
\\&-\ket{0}_{1}\ket{1}_{2}\ket{1}_{3}).
\end{align}
This state is a three-particle Greenberger-Horne-Zeilinger-class
state (see, for example, \cite{Rubens-ent-class}). It has highest
(unit) ``three-tangle'' or ``residual entanglement'' (proposed by
Coffman, Kundu and Wooters \cite{Wooters-e-part-ent-measure}): the
concurrence \cite{concurrence-1,concurrence-2} of each qubit with
the rest of the system is $1$, and all the pairwise concurrences are
$0$. A three-particle GHZ-class state is also obtained from Eq.
(\ref{ent-state-gen-4}).

\par
\emph{Controlling the Amount of Entanglement}.\textemdash In an
actual experiment, the path identity can be partially (or fully)
lost. Importantly, the loss of path identity can be controlled by
inserting an attenuator (neutral density filter for photons) in the
path of aligned particles between the two sources. We now analyze
such a situation and show that it is possible to control the amount
of entanglement without interacting with the entangled particles.
\par
We consider the general scheme (Fig. \ref{fig:gen-setup}) and in
addition we assume that attenuators are placed between $Q$ and $Q'$
in each of the beams $b_l$, where $l=N-M+1,\dots,N$. The quantum
state generated by the two $N$-particle sources is again given by
Eq. (\ref{state-N-part}). However, the transformation of the states
due to alignment of particle paths is now given by
\cite{Note-qft-sup}
\begin{equation}\label{alignment-cond-gen}
\ket{b_l}_l \to
\exp[i\theta_l]~(T_l\ket{b_l'}_l+\sqrt{1-T_l^2}\ket{v}_l),
\end{equation}
where $0\leq T_l \leq 1$ is the amplitude transmission coefficient
of an attenuator ($1-T_l^2$ is the probability of particle $l$
getting lost before arriving at $Q'$), $\ket{v}_l$ represents the
state of a lost particle, and $l=N-M+1,\dots,N$. Clearly, $T_l=1$
implies no loss of path identity (for particle $l$) and $T_l=0$
implies complete loss of path identity.
\begin{figure}[b]  \centering
\includegraphics[width=0.4\textwidth]{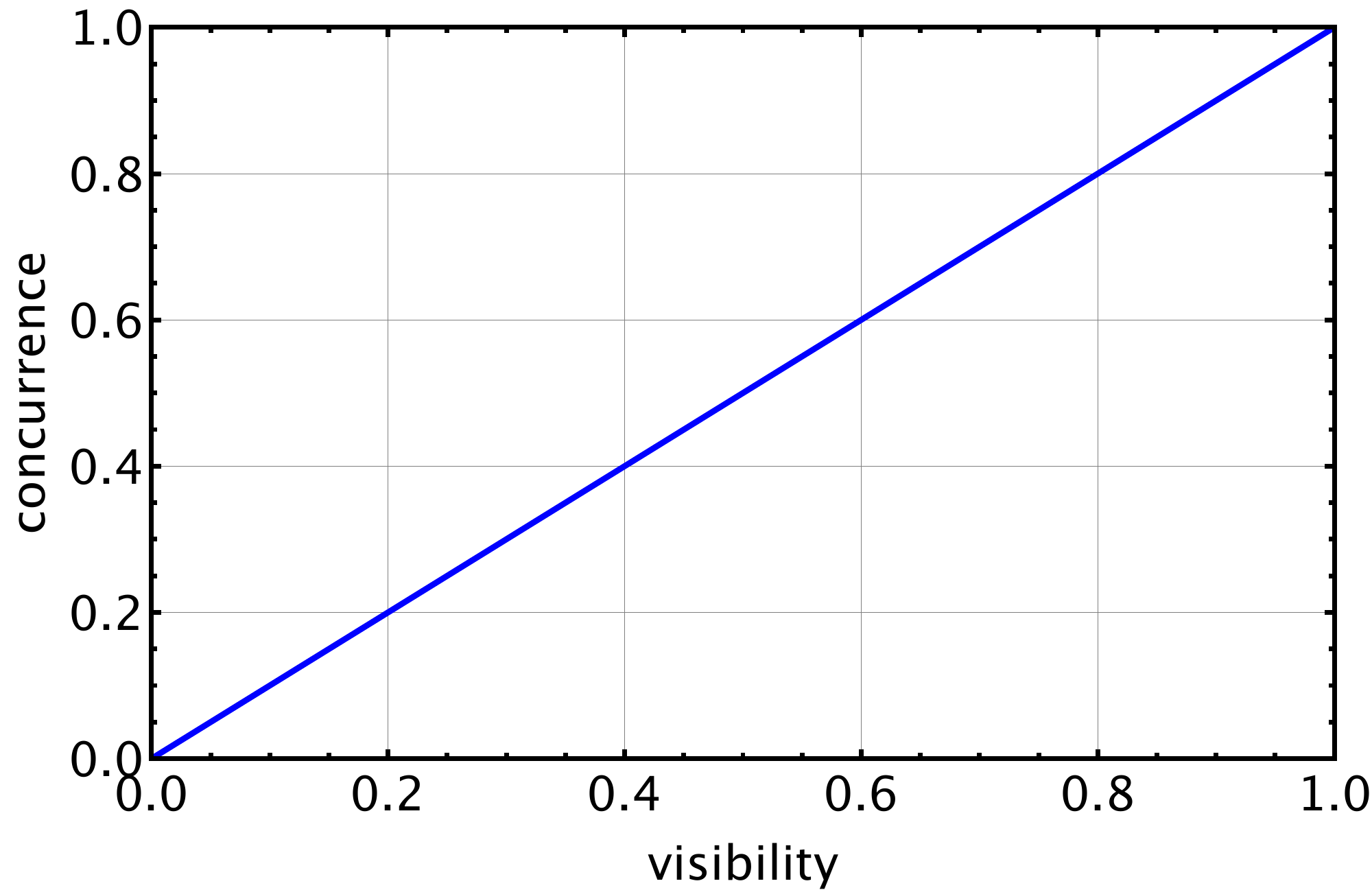}
\qquad \caption{Controlling the amount of entanglement. Two-particle
entangled states are produced using the setup illustrated by Fig.
\ref{figa:3-part-1-Pi-setup}. The concurrence is equal to the
visibility of the two-particle interference pattern. Both
concurrence and visibility are equal to the amplitude transmission
coefficient of the attenuator (when there is no experimental loss).}
\label{figa:concplot}
\end{figure}
\par
The transformations of the states due to beam splitters are given by
Eq. (\ref{BS-1}), where $j=1,2,\dots,N-M$. The many-particle
interference patterns and the many-particle entangled states are
obtained by applying Eqs. (\ref{BS-1}), (\ref{state-N-part}), and
(\ref{alignment-cond-gen}). It is to be noted that the particles
emerging from the beam splitters are in a mixed state when $T_l\neq
1$ for any $l$. The density operator representing this state is
obtained by taking partial trace over the undetected modes and the
loss modes. Below we illustrate the method by an example.
\par
Let us consider the situation illustrated by Fig. (2a) with the
additional assumption that an attenuator is placed in beam $b_3$
between $Q$ and $Q'$. In this case, $N=3$ and $M=1$. Applying Eqs.
(\ref{BS-1}), (\ref{state-N-part}), and (\ref{alignment-cond-gen}),
we find that
\begin{align}\label{state-thr-part-final}
\ket{\psi_0} \to
&\ket{\psi}=\frac{1}{2}\left[(T_3-e^{i\zeta^{(3)}_1})\ket{b'_3}_3
+\sqrt{1-T_3^2}\ket{v}_3\right]\ket{\Phi^{-}} \nonumber
\\&+\frac{i}{2}\left[(T_3+e^{i\zeta^{(3)}_1})\ket{b'_3}_3
+\sqrt{1-T_3^2}\ket{v}_3\right]\ket{\Psi^{+}},
\end{align}
where $\zeta^{(3)}_1=\phi_0+\phi_1+\phi_2-\theta_3$; and
$\ket{\Psi^{+}}$ and $\ket{\Phi^{-}}$ are given by Eq.
(\ref{Bell-st-output}). The density operator, $\widehat{\rho}$,
representing the quantum state of the particles emerging from the
beam splitters is obtained by taking the partial trace of
$\ket{\psi}\bra{\psi}$ over $\ket{b'_3}_3$ and $\ket{v}_3$. We thus
have
\begin{align}\label{do-form}
\widehat{\rho}=&\text{tr}\left\{\ket{\psi}\bra{\psi}\right\}_{b_3',v}
=\frac{1}{2}\left(1-T_3\cos\zeta^{(3)}_1 \right)
\ket{\Phi^{-}}\bra{\Phi^{-}} \nonumber
\\& \quad +\frac{1}{2}
\left(1+T_3\cos\zeta^{(3)}_1\right)\ket{\Psi^{+}}\bra{\Psi^{+}}
\end{align}
\par
It follows from Eqs. (\ref{Bell-st-output}) and (\ref{do-form}) that
the rate of coincidence detection rate of particles $1$ and $2$ at
the pairs of detectors ($d_1$,$d_2$), ($d_1$,$d_2'$),
($d_1'$,$d_2$), and ($d_1'$,$d_2'$) are given by
\begin{subequations}\label{two-part-pattern-loss}
\begin{align}
P_{d_1d_2}=P_{d_1'd_2'}&=\frac{1}{4}[1-T_3\cos\zeta^{(3)}_1], \label{two-part-pattern:a} \\
P_{d_1d_2'}=P_{d_1'd_2}&=\frac{1}{4}[1+T_3\cos\zeta^{(3)}_1].
\label{two-part-pattern:b}
\end{align}
\end{subequations}
These two-particle interference patterns are similar to the ones
given by Eq. (\ref{two-part-pattern}), except they no longer have
unit visibility. The visibility is now given by
\begin{align}\label{vis-form}
\mathcal{V}=T_3.
\end{align}
\par
If we choose $\zeta^{(3)}_1=2m\pi$, Eq. (\ref{do-form}) reduces to
\begin{align}\label{do-form-sp-1}
\widehat{\rho}_{\text{even}}=\frac{1}{2}\left(1-T_3 \right)
\ket{\Phi^{-}}\bra{\Phi^{-}}+\frac{1}{2}
\left(1+T_3\right)\ket{\Psi^{+}}\bra{\Psi^{+}},
\end{align}
and for $\zeta^{(3)}_1=(2m+1)\pi$, we get
\begin{align}\label{do-form-sp-2}
\widehat{\rho}_{\text{odd}}=\frac{1}{2}\left(1+T_3 \right)
\ket{\Phi^{-}}\bra{\Phi^{-}}+\frac{1}{2}
\left(1-T_3\right)\ket{\Psi^{+}}\bra{\Psi^{+}}.
\end{align}
Clearly, when the coincidence detection rates at ($d_1$,$d_2'$) and
($d_1'$,$d_2$) maximize, the state given by Eq. (\ref{do-form-sp-1})
is obtained. Similarly, when the coincidence detection rates at
($d_1$,$d_2$) and ($d_1'$,$d_2'$) maximize, the state given by Eq.
(\ref{do-form-sp-2}) is obtained.
\par
We now investigate the amount of entanglement in these mixed states.
For simplicity of notation we represent the unprimed state by $0$
and primed states by $1$. In this notation, we have
$\ket{d_1}_1\ket{d_2}_2\equiv \ket{0,0}$,
$\ket{d_1}_1\ket{d_2'}_2\equiv \ket{0,1}$,
$\ket{d_1'}_1\ket{d_2}_2\equiv \ket{1,0}$, and
$\ket{d_1'}_1\ket{d_2'}_2\equiv \ket{1,1}$. In this basis, the mixed
states given by Eqs. (\ref{do-form-sp-1}) and (\ref{do-form-sp-2})
take the following matrix forms:
\begin{subequations}\label{do-form-mat}
\begin{align}
&\left[\widehat{\rho}_{\text{even}}\right]=\left(
 \begin{array}{cccc}
   \frac{1-T_3}{4} & 0 & 0 & -\frac{1-T_3}{4} \\[6pt]
   0 & \frac{1+T_3}{4} & \frac{1+T_3}{4} & 0 \\[6pt]
   0 & \frac{1+T_3}{4} & \frac{1+T_3}{4} & 0 \\[6pt]
   -\frac{1-T_3}{4} & 0 & 0 & \frac{1-T_3}{4}
  \end{array} \right),
\label{do-form-mat:a}
\\[6pt]
&\left[\widehat{\rho}_{\text{odd}}\right]=\left(
 \begin{array}{cccc}
   \frac{1+T_3}{4} & 0 & 0 & -\frac{1+T_3}{4} \\[6pt]
   0 & \frac{1-T_3}{4} & \frac{1-T_3}{4} & 0 \\[6pt]
   0 & \frac{1-T_3}{4} & \frac{1-T_3}{4} & 0 \\[6pt]
   -\frac{1+T_3}{4} & 0 & 0 & \frac{1+T_3}{4}
  \end{array} \right).
\label{do-form-mat:b}
\end{align}
\end{subequations}
We determine the concurrence using the standard procedure
\cite{concurrence-2} and find that both states have the same
concurrence
\begin{align}\label{conc}
\mathcal{C}(\widehat{\rho})=T_3.
\end{align}
Comparing Eqs. (\ref{vis-form}) and (\ref{conc}), it becomes clear
that
\begin{align}\label{conc-vis}
\mathcal{C}(\widehat{\rho})=\mathcal{V},
\end{align}
i.e., in this case the concurrence is equal to the visibility of the
two-particle interference pattern [Fig. \ref{figa:concplot}]. We
note that one can change both the concurrence and the visibility by
varying $T_3$. Since the attenuator never interacts with the
entangled particles, \emph{the scheme allows us to control the
amount of entanglement in an interaction-free way}.
\par
The method also applies when the number of entangled particles is
more than two. This is because for any number of particles, the
placement of the attenuators results in the conversion of a pure
output state to a mixed one.
\begin{figure}[htbp]  \centering
\includegraphics[width=0.4\textwidth]{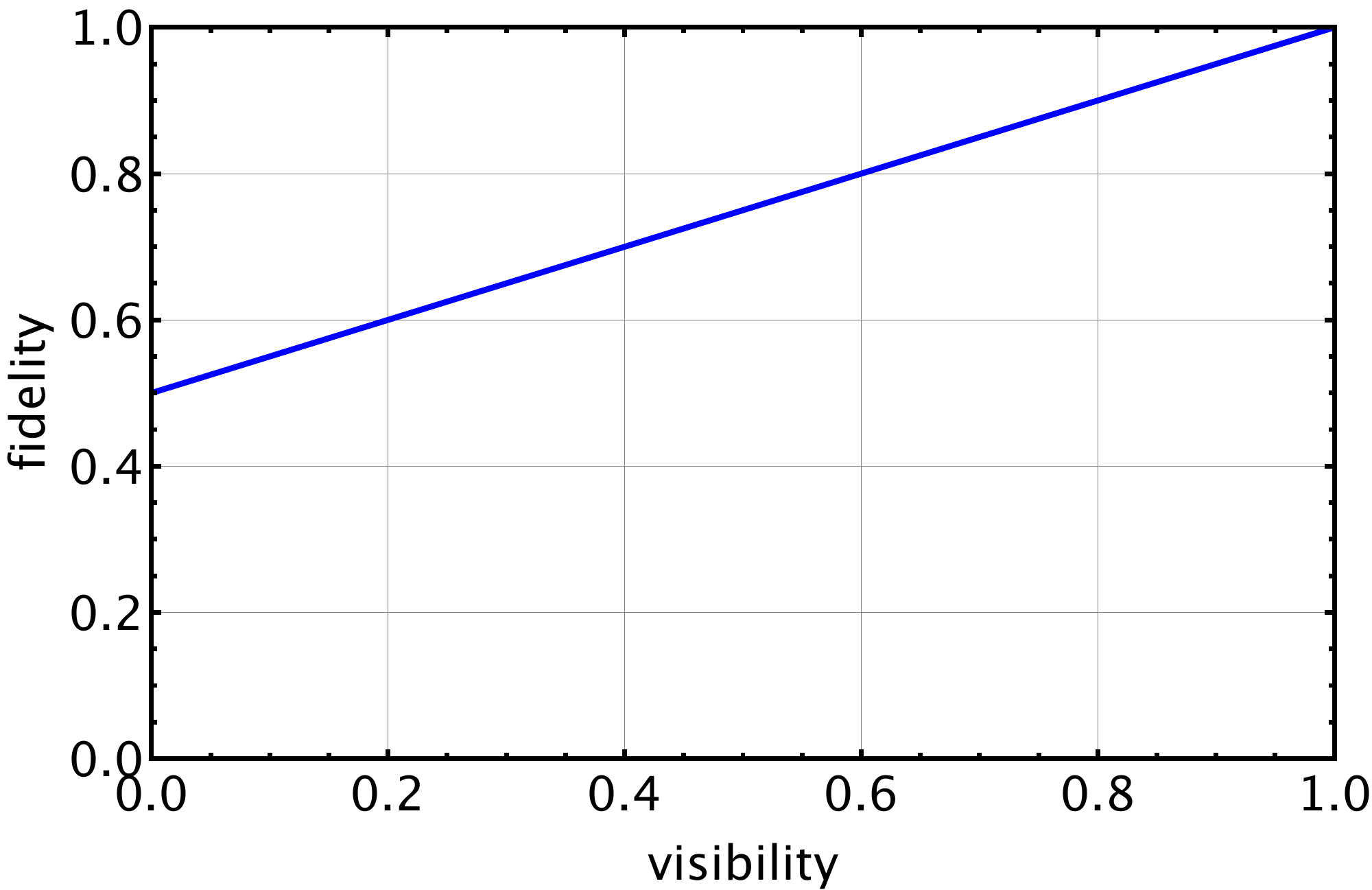}
\qquad \caption{Fidelity and loss of path identity. We consider the
case illustrated by Fig. \ref{figa:3-part-1-Pi-setup}. The fidelity
varies linearly with the visibility of the two-particle interference
pattern. No path identity results in zero visibility and perfect
path identity results in highest possible visibility.}
\label{figb:fidplot}
\end{figure}
\par
\emph{Fidelity}.\textemdash Finally, we briefly discuss the fidelity
for the output states. Determining fidelity is relevant when the
loss of path identity is unintended and due to experimental
imperfections. Equation (\ref{alignment-cond-gen}) again applies in
this case but $T_l$ now signifies the quality of alignment or other
experimental losses.
\par
Let us once again consider the situation illustrated by Fig. (2a).
Without any experimental imperfections, one would expect the output
state to be $\ket{\Psi^{+}}$ for $\zeta^{(3)}_1=2m\pi$. However,
when there is a loss of path identity, the output state is
represented by $\widehat{\rho}_{\text{even}}$ [Eq.
(\ref{do-form-sp-1})]. The fidelity is, therefore, given by
\cite{fidelity-Jozsa}
\begin{align}\label{fidelity-1}
\mathcal{F}\equiv \bra{\Psi^{+}}\widehat{\rho}_{\text{even}}
\ket{\Psi^{+}}=\frac{1}{2}(1+\mathcal{V}),
\end{align}
where we have used Eq. (\ref{vis-form}). Similarly, for
$\zeta^{(3)}_1=(2m+1)\pi$, the fidelity is again equal to
$(1+\mathcal{V})/2$. Clearly, the fidelity is directly related to
the visibility of the two-particle interference pattern [Fig.
\ref{figb:fidplot}].
\par
The loss of path identity results in the conversion of a pure output
state into a mixed one for any number of particles. We therefore
expect that a relationship between fidelity and visibility also
exists when the number of particle increases.
\par
\emph{Conclusions}.\textemdash We have introduced a novel scheme of
many-particle interferometry that can be used for producing
many-particle entangled states. In contrast to a series of notable
studies (see, for example,
\cite{GHZ-original,Z-three-part-interf,HZ-EPR-interf,Zukowski-Bell-th-no-pol,Z-two-part-interf,Rarity-Tapster-Bell-no-pol,Pan-Z-Zuk-RMP})
that have already emphasized the connection between entanglement and
interference, our work uses the concept of path identity.
\par
In our scheme, path identity is a result of the fact that both
sources can emit a certain number ($M$) of particles into the same
modes of the associated quantum field. Therefore, the scheme is
applicable to any quantum system (e.g., atoms, fundamental
particles) that can be treated in the framework of quantum field
theory \cite{Note-Raman-Mandel}.
\par
Our scheme produces many-particle entangled states that are
superpositions of different Dicke states. We have also shown that
using this scheme, maximally entangled two-qubit states (Bell
states) and maximally three-tangled quantum states
(Greenberger-Horne-Zeilinger-class states) can be produced. We
expect that further investigations regarding the states produced by
our scheme will lead to promising results.
\par
An important feature of our scheme is that the generated entangled
states can be manipulated without interacting with the entangled
particles. Furthermore, the scheme also allows us to control the
amount of entanglement in a quantum state. We hope that this type of
quantum state control and engineering will have a significant impact
in quantum information science.
\par
Finally, our scheme can be further generalized by including other
degrees of freedom, for example, polarization, orbital angular
momentum, etc., for the photonic cases. Another generalization will
be the use of multiport beam splitters \cite{Reck-Z-multiport}
instead of the standard two-port beam splitters. It will also be
interesting to investigate whether our scheme can be represented and
further analyzed by the graph theoretical technique that has
recently been introduced by Krenn, Gu, and Zeilinger
\cite{Krenn-qexp-graph}.
\par
\emph{Acknowledgements}.\textemdash The author thanks Professor
Anton Zeilinger for his careful reading of the manuscript and his
numerous valuable comments. The author also thanks Mr. Manuel
Erhard, Dr. Mario Krenn, and Professor K. Birgitta Whaley for
fruitful discussions. This work was supported by the Austrian
Academy of Sciences (\"OAW- IQOQI, Vienna), and the Austrian Science
Fund (FWF) with SFB F40 (FOQUS).


\begin{thebibliography}{99}
\bibitem {Note-q-sup} If one cannot determine (even in principle)
from which source the particles are emitted, the resulting state is
in a pure quantum superposition.
\bibitem {Note-qft} A fully equivalent quantum state is obtained if
one employ quantum field theory. Treatment of the beam alignment
using quatized fields can be found in Refs.
\cite{ZWM-ind-coh-PRL,WZM-ind-coh-PRA,LLLZ-th-mandel-img}.
\bibitem {Note-spin-free-states} Such spin-free two-particle
entangled states have been discussed before in the context of
traditional two-particle interferometry and violation of Bell's
inequality (see, for example,
\cite{HZ-EPR-interf,Zukowski-Bell-th-no-pol,Z-two-part-interf,Rarity-Tapster-Bell-no-pol}).
\bibitem {Note-qft-sup} This treatment is equivalent to the treatment
based on quantum field theory given in Ref. \cite{ZWM-ind-coh-PRL}.
\bibitem {Note-Raman-Mandel} Ref. \cite{Ou-Raman-Mandel} demonstrated that two atomic
ensembles can be made indistinguishable by employing the path
identity of Stokes photons created by Raman scattering. It also
suggested that the use of many atomic ensembes can, in principle,
make them path-entangled.
\end{thebibliography}
\end{document}